\documentclass[preprint,sort&compress]{elsarticle}

\usepackage[T1]{fontenc}
\usepackage[english]{babel}
\usepackage{subfig}
\usepackage{hyperref}

\journal{Solid State Ionics}

\begin{document}

\begin{frontmatter}

\title{Numerical modeling of the heterocycle intercalated
proton-conducting polymers at various mole ratios}
\author[md]{T.~Mas\l{}owski\corref{cr}}
\ead{T.Maslowski@if.uz.zgora.pl}
\author[md]{A.~Drzewi\'{n}ski}
\author[pl]{P.~\L{}awniczak}
\author[ju]{J.~Ulner}
\cortext[cr]{Corresponding author}
\address[md]{Institute of Physics, University of Zielona G\'{o}ra,
ul. Prof. Z. Szafrana 4a, 65-516~Zielona~G\'{o}ra, Poland}
\address[pl]{Institute of Molecular Physics, Polish Academy of Sciences,
ul. M. Smoluchowskiego 17, 60-179~Pozna\'{n}, Poland}
\address[ju]{Institute of Low Temperature and Structure Research PAN, 
ul. Ok\'{o}lna 2, 50-422~Wroc\l{}aw, Poland}

\date{\today}

\begin{abstract}
The kinetic Monte Carlo simulations are employed to study the proton 
conductivity for anhydrous heterocyclic based polymers. The proton transport 
is based on a two-step process called the Grotthuss mechanism. In the 
referring system the proton concentration depends on the relative molar ratio, 
$x$, of the benzimidazole and the polystyrene sulfonic acid. Available 
experimental data with contrasting behavior are fitted and interpreted in 
terms of our microscopic model. Moreover, it has been shown that the current 
behavior similar to the Vogel-Tamman-Fulcher law can be reproduced with high 
precision on the basis of the Grotthuss mechanism.
\end{abstract}


\end{frontmatter}

\section{Introduction}\label{S1}

The microscopic modeling of proton transport is one of the long-standing 
problems in many areas of science ranging from conversion of chemical energy 
into electrical one to various biological systems \cite{Colomban,Kreuer_2}. 
Ice and water became the first hydrogen-bonding systems for which the 
microscopic description of proton defect transport has been provided in great 
detail \cite{Schuster}. However, as there are several advantages of fuel cells 
operating above the boiling temperature of water \cite{Honma}, recently the 
polymer systems which conduct protons in the absence of any water have become 
the subject of intensive research. Unfortunately, the proton conductivity of 
conventional polymer membranes under anhydrous condition is usually very low. 
Therefore, the promising strategy for the synthesis of new materials was the 
doping of a high boiling proton solvent into a polymer matrix. Then the proton 
transport occurs almost entirely through the Grotthuss mechanism, a two-stage 
mechanism \cite{Grot,Kreuer}, consisting of thermally induced reorientation 
and proton tunneling in hydrogen bonds (H-bonds).

As is well known, both chemical and physical properties of a polymer material 
may change with substitution. So to get polymer electrolytes the polymer 
matrix (e.g., polystyrene, polyacrylate, polysiloxane) is doped by amphoteric 
nitrogen-based heterocycles (e.g., imidazole, triazole, benzimidazole) at 
various mole ratios \cite{Kreuer_0,Kreuer_1,Celik}. As a result the 
heterocycles may be covalently tethered to a suitable polymer and linked by 
the N--H$\cdots$N hydrogen bridges providing a migration path for excess of 
protons emerging from the dissociation of the acid functions.

For many heterocyclic based polymer proton conductors the temperature 
dependence of dc conductivity follows at low doping ratios $x \leq 1.0$ ($x$ 
is the number of moles of heterocycle per polymer repeat unit containing the 
acidic group) the simple Arrhenius law 
\begin{equation} \label{arr}
\sigma = \sigma_{0} \exp \left( E_a/k_B T \right),
\end{equation}
where $E_a$ is the activation energy for proton migration, $k_B$ is Boltzmann's 
constant and $\sigma_0$ corresponds to carrier proton number. But what is 
intriguing, at a high doping the temperature-dependent conductivity seems to 
follow the Vogel-Tamman-Fulcher (VTF) law \cite{Bozkurt,exper,Vila}
\begin{equation} \label{vtf}
\sigma = \sigma_{0} \exp \left( E_a/k_B (T-T_0) \right).
\end{equation}
typically associated with a viscous material whose conductivity is driven by 
the segmental motions above the glass transition temperature. The parameter 
$T_0$ corresponds to the temperature where the free volume disappears. 
Moreover, for these compounds the crossover driven by the molar ratio is 
accompanied by an abrupt increase in proton conductivity for $1<x<1.5$.

To address this controversy, we have evaluated the experimental data in the 
wide range of doping ratio \cite{Bozkurt} employing the model \cite{PRE} based 
on the kinetic Monte Carlo (KMC) simulations \cite{KMC,You66,Fic91,Hermet} 
which we describe briefly here. The proton conduction process of the 
immobilized heterocycles can be considered as a cooperative one involving both 
molecular motions prior to the proton exchange (by the 180$^{\circ}$ flip) and 
migration along the H-bond chain. It can be well modeled by the 
one-dimensional system of rods each of which has only two positions where the 
rod ends can be occupied by protons. The key point is to know {\it a priori} 
all transition rates from every configuration to any other allowed one 
\cite{Fic91}. When it is satisfied the KMC method gives the answer to the 
question of how long the system remains in the same configuration and to what 
configuration it will evolve \cite{KMC}. Herein, the rotations, as a 
sub-process, are treated as a thermally activated process satisfying the 
Arrhenius law [Eq.~(\ref{arr})] with the activation energy, $E_a$, given by
\begin{equation}
E_a=\max\left(0,V_\mathrm{act}+|e|Kb\right) \;,
\end{equation}
where $V_\mathrm{act}$ is the activation energy for rotation in the vanishing 
electric field, $e$---the value of the elementary charge, $K$---the external 
electric field strength, and $b$---the size of the rod. The prefactor 
$\sigma_{0}$ (in this case called the frequency of rotation $\nu^0_R$) is 
calculated by solving the Schr\"{o}dinger equation for the one-dimensional 
quantum rotor (see \cite{PRE}).

The migration of a proton from one rod to another represents the hopping 
between the minima of the H-bond potential. Hopping is defined as the 
thermally assisted tunneling which is an extension of the purely classical 
Arrhenius behavior. We approximate the H-bond potential by the fuzzy Morse 
potentials originating in rod ends as they represent anionic groups between 
which the H-bonds are created.

\begin{eqnarray}
V_a(z)&=&\frac{1}{2a}\int_{-a}^a \left[V_{\rm Morse}\left(\frac{d}{2}-z+y\right)
\right. \nonumber \\
&&\left. +\, V_{\rm Morse}\left(z-y-\frac{d}{2}\right)\right] d y\;,\label{va}\\
V_{\rm Morse}(z) &=& g \left[\exp\left(-\frac{2z}{w}\right) 
-2 \exp\left(-\frac{z}{w}\right)\right]\;. \label{pot}
\end{eqnarray}

$V_a(z)$ is the double well potential and the parameter $a$ controls the 
dispersion in the position of the anionic groups forming the H-bond. Thus, it 
represents the thermal lattice vibrations. The Morse potential parameters $g$ 
and $w$ are adjusted to get the distance between the minima of the double well 
potential $V_a$ equal to $\Delta z$ together with the height of the barrier 
equal to $h$. $V_a(z)$ is used to get quantum analog of the Arrhenius law 
\cite{Bell}. In this case the prefactor $\sigma_{0}$, called the frequency of 
tunneling, is denoted by $\nu^0_T$. More details about the model can be found 
in our previous paper \cite{PRE}.

This paper is organized as follows: in Sec.~\ref{S2} the proton concentration 
with respect to the molar ratio is determined. In Sec.~\ref{S3} the referring 
experimental system is presented as well as the simulation results are 
discussed. Section~\ref{S4} concludes our paper summarizing the main outcomes.

\section{Molar ratio influence on the proton concentration}\label{S2}

\begin{figure}[htb]
\centering
\includegraphics[width=78mm]{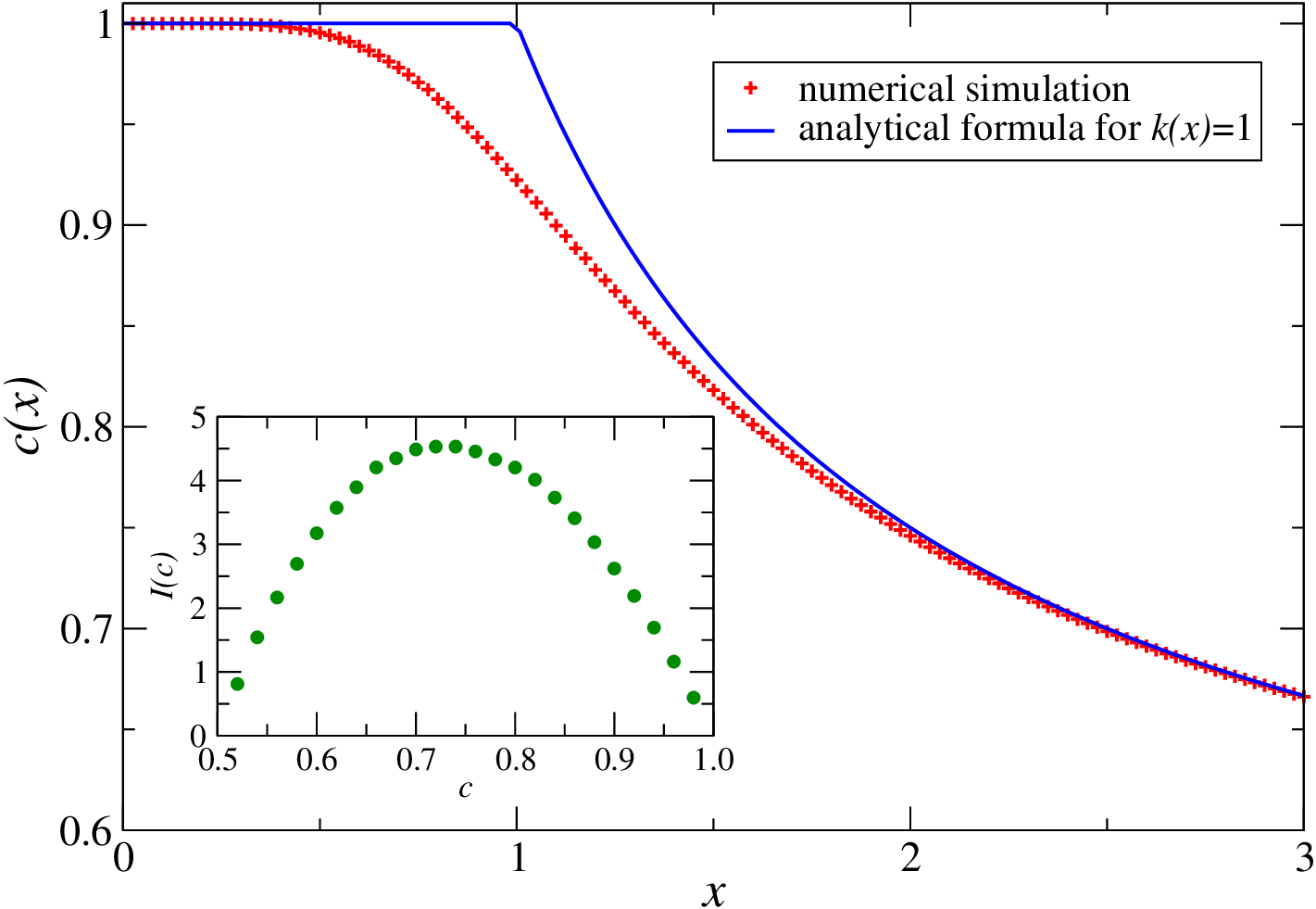}
\caption{The proton concentration as the function of the molar ratio, the 
result for the $125\times125\times125$ system. The solid line represents the 
asymptotic behavior. Inset: The current dependence on the proton concentration 
\cite{PRE}.}
\label{plot1}
\end{figure}

The blends of polymer and heterocycles are prepared for various $x$ where the 
result are transparent and homogeneous thin films. The larger ratio $x$ the 
lower the average distance between the heterocycles, which in turn 
significantly modifies the character of the H-bond potential. As the 
amphoteric nitrogen-based heterocycles demonstrate the presence of both 
protonated and non-protonated nitrogen atoms they can act as donors and 
acceptors in proton-transfer reactions. Therefore, the change of $x$ affects 
the concentration of protons, $c$, on the heterocycle path: a parameter 
crucial for the proton conductivity (\cite{PRE}, see the inset in 
Fig.~\ref{plot1}).

In order to determine the function $c(x)$ we have performed the numerical 
simulations considering two types of particles, $A$ and $B$ with the relative 
ratio $x$, randomly distributed at sites on a simple cubic lattice. When the 
adjacent sites are occupied by $A$ and $B$, they may form a pair (one particle 
can be paired only once). The concentration $c(x)$ may be expressed in terms 
of the function $k(x)$ describing the probability that $A$ is attached to $B$:
\begin{equation}
c(x)=\frac{1}{2}\left[1+k(x)\right]\;.
\end{equation}
The function $k(x)$ is calculated as the ratio of paired $A$ particles to the 
number of all $A$ particles. Simulations were performed for system sizes 
$75\times75\times75$, $100\times100\times100$ and $125\times125\times125$. 
The function $c(x)$ converges quickly with the system size, providing that the 
differences between the results for the two largest systems are 
indistinguishable.

If a simplifying assumption is considered that $A$ is always attached to $B$, 
provided that there is a free $B$ particle then simple analytical expressions 
for asymptotes can be derived: $c(x)=1$ when $x$ goes to zero and 
$c(x)=(x+1)/(2x)$ when $x$ is large (see Fig.~\ref{plot1}). 

In our case $A$ represents a heterocycle while $B$ a polymer unit, and pairing 
should be understood as forming the anionic group by deprotonation of the 
acidic group by doping with the heterocycle, e.g., $\mathrm{BnIm} + 
\mathrm{SO}_{3}\mathrm{H} \rightarrow \mathrm{BnImH}^{+} + 
\mathrm{SO}_{3}^{-}$ (the benzimidazole is abbreviated to BnIm). In this way 
some heterocycles are protonated at "free" nitrogen site and the proton 
concentration increases. The fall of the curve for the increasing $x$ is 
associated with the relative excess of heterocycles which can hardly come 
across any free acidic group.

\section{Benzimidazole intercalated polymer}\label{S3}

In this section the available experimental data for the benzimidazole-based 
polymer proton conductors \cite{Bozkurt} are analyzed. For the pure 
polystyrene the glass-transition temperature is $T_g=95^{\circ}{\rm C}$ while 
for the polystyrene sulfonic acid (PSSA) is shifted to $T_g=140^{\circ}{\rm C}$ 
\cite{Martins}. After intercalation of the PSSA within the BnIm, the $T_g$ 
rises only slightly to $149^{\circ}{\rm C}$, hardly depending on the molar 
ratio $x$. When for the same polymer matrix the imidazole or triazole are 
dopants they act as plasticizers shifting the $T_g$ values to lower 
temperatures \cite{Bozkurt2}. Moreover, in such a case the $T_g$ depends 
significantly on the molar ratio.

The BnIm is anchored to the polymer backbone by the covalent bond. The FT-IR 
spectra show that the SO$_{3}$H groups are deprotonated by doping with the 
benzimidazole and form SO$_{3}^{-}$ groups \cite{Bozkurt}. This in turn shall 
increase the concentration of protons traveling along the conduction pathways 
with respect to the pure BnIm (one proton per one BnIm molecule). But since 
the highest current was measured at $x=1.5$, it confirms the well-known fact 
that the intensive diffusion requires both a high charge density and a high 
defect density. Note that when the relative molar ratio grows, more and more 
heterocycles remain not associated with the acidic groups. They are 
functioning as an insertion into the polymer structure. Because they are not 
protonated at "free" nitrogen site, such a site provides a defect necessary to 
have the efficient Grotthuss mechanism.

In our simulations, in general, the parameters have been derived from fitting 
the experimental data for the proton current at $x=1.5$ (see Appendix). It 
should be emphasized that the interplay between the thermal expansion and the 
thermal vibration \cite{PRE} is crucial to obtain a high accuracy curve 
fitting. The parameters of the H-bond potential found for $x=1.5$ were applied 
for $x=1.0$ and $0.5$. But when $x$ increases the average distance between 
benzimidazoles becomes smaller and the activation energy for thermal hopping 
decreases as well. Similarly, one can expect that the activation energy for 
rotation of such an unbound heterocycle is lower than that of the immobilized 
one. So, the effective activation barrier for rotation is likewise to decrease 
with increasing $x$. It is also reasonable to expect that when the temperature 
rises the overall rotational barrier is reduced. Unfortunately, the function 
describing the change of the rotational potential with increasing temperature 
is unknown. Therefore, we decided to carry out the simulations for the 
simplest case when the value of the rotational barrier is independent of 
temperature.

\begin{figure}[htb]
\centering
\subfloat{
\includegraphics[width=78mm]{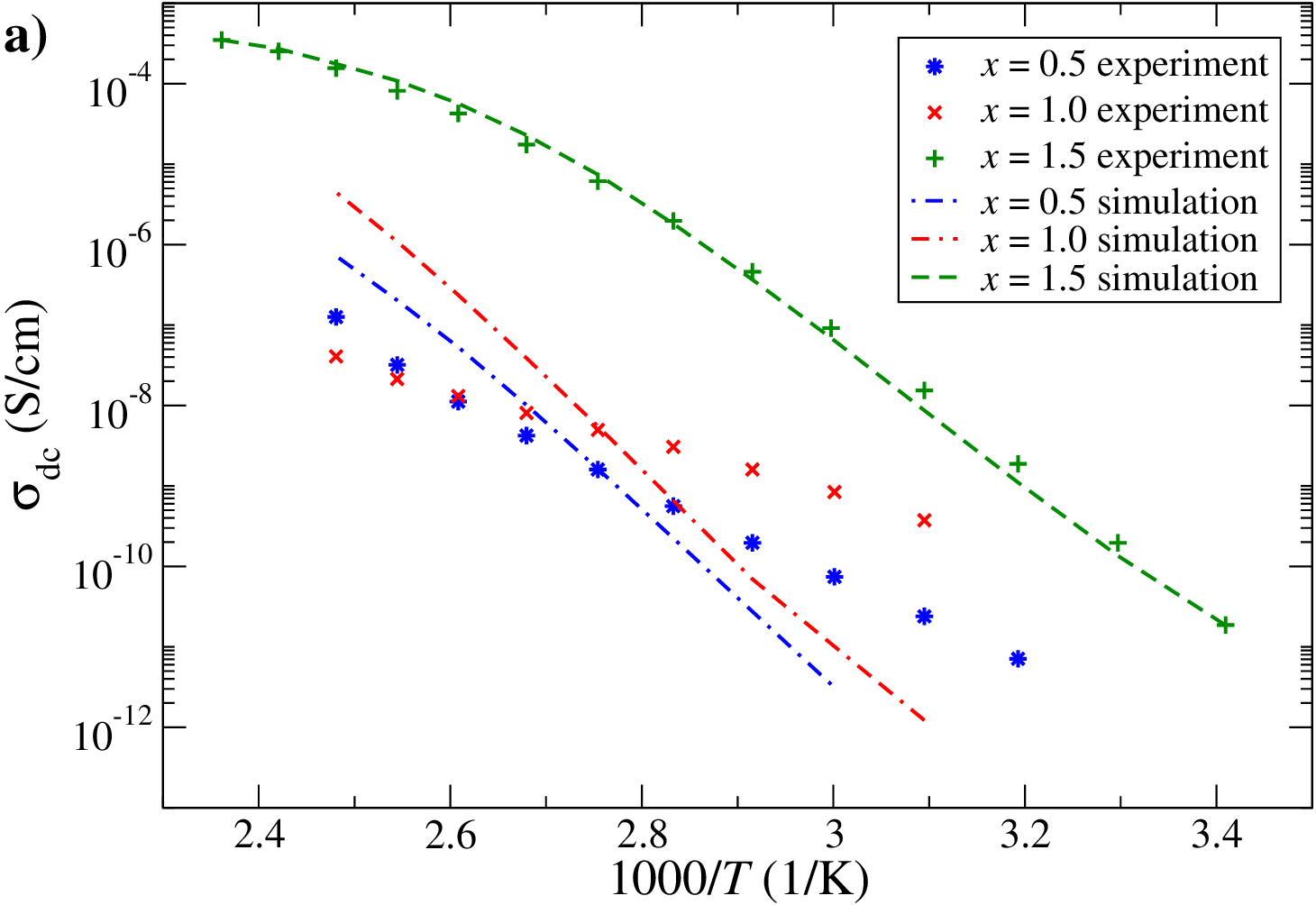}
\label{plot2a}
}\\
\bigskip
\bigskip
\subfloat{
\includegraphics[width=78mm]{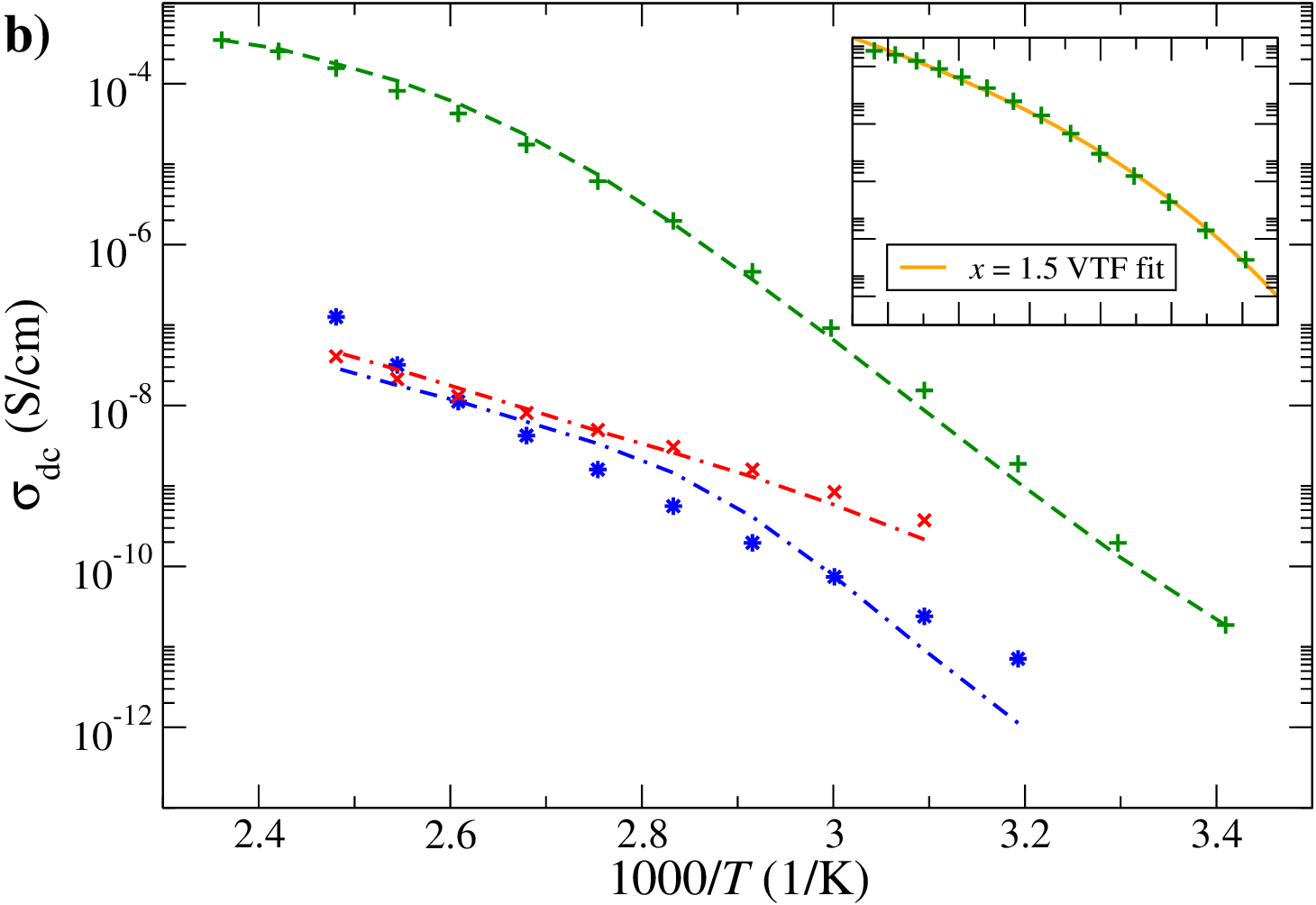}
\label{plot2b}
}
\caption{The dashed lines represent the simulation fits, whereas the symbols 
show the experimental data: a) only the H-bond distance is modified to obtain 
the best fit, b) both the H-bond distance and the activation energy for 
rotation are chosen to get the best fit. Specific parameter values are 
provided in the Appendix. Inset: The best VTF fit to the $x=1.5$ experimental 
data. The same scales on the axes as in the main figure.}
\end{figure}

To verify the role of the activation energy for rotation, our simulations were 
performed in two ways, namely, the activation energy for rotation was fixed 
and only the H-bond distance was decreasing with increasing $x$ 
(Fig.~\ref{plot2a}) or it was depended on the molar ratio as well as the 
H-bond distance (Fig.~\ref{plot2b}). The agreement with the experiment is 
considerably better for the second case.

As one can see in Fig.~\ref{plot2b} our microscopic model is able to cover 
both the Arrhenius behavior for $x=1$ as well as the VTF-like behavior for 
$x=1.5$. It should be emphasized that below the glass transition the current 
behavior similar to the VTF law can be reproduced with high precision on the 
basis of the Grotthuss mechanism. Moreover, the best VTF fit in accordance 
with Eq.~(\ref{vtf}) exhibits $T_0=-75$ $^{\circ}{\rm C}$ in contrast to the 
experimental value of the glass transition $T_g=149$ $^{\circ}{\rm C}$ 
determined by the differential scanning calorimetry. $T_0$ is the temperature 
at which the segmental transport ceases to exist and it is extremely hard to 
believe that this takes place so far below the glass transition \cite{Woud}. 
Hence, the VTF-like shape of the conductivity curve does not mean that we have 
to deal with the proton transport with the assistance of polymer chain 
segmental motion.
\bigskip
\bigskip

The numerical results for $x = 0.5$ are moved away from the expected 
measurement values. This discrepancy is likely to be related to the low 
concentration of heterocycles making that an unbroken path for the proton 
transport is formed very rarely. In the percolation terminology it seems to be 
associated with the attendance of the percolation threshold for $0.5 < x < 1$. 
One of the assumptions of our one-dimensional model is the presence of a 
significant number of conductive paths. Thus, the range of the low molar 
ratios appear to be beyond the range of applicability of the model.

\section{Conclusion}\label{S4}
The BnIm intercalated in the PSSA matrix can be employed as the polymer 
electrolyte membrane holding promise for the synthesis of more temperature 
tolerant proton conductive membranes. Provided our microscopic approach takes 
into account the influence of the mole ratio both on the average distance 
between benzimidazoles, proton concentration and activation energy for 
rotation we are able to reproduce both measurements following the Arrhenius 
law and the VTF-like law. Moreover, an abrupt increase in proton conductivity 
for $1<x<1.5$ can be accurately reproduced.

Simulation results for the highest molar ratio ($x = 1.5$) show that the 
current behavior similar to the VTF law should not always be attributed to 
segmental motion of the polymer matrix. Sometimes, below the glass transition, 
a subtle interplay of parameters governing the proton transport can lead to 
the VTF-like behavior despite the fact that we are dealing only with the 
Grotthuss mechanism.

As for the polycrystalline benzimidazole the current studies \cite{PRE} have 
demonstrated that the thermal lattice vibrations, which modify the H-bond 
potential, play an essential role in the conduction process. Overall, the 
validity of the model has been successfully examined for two different systems 
based on the BnIm. So, it can be concluded that the major diffusion processes 
have been correctly captured in our model formulation.

\section*{Acknowledgments}
This work was supported by the Polish Ministry of Science and Higher Education 
through Grant No. N N202 368139. We have benefited from discussions with Maria 
Zdanowska-Fr\k{a}czek.

\appendix

\section{Model parameters}
$d_0$ and $a_0$ are the values of $d$ and $a$ at $T=T_0$, while $d_1$ and 
$a_1$ are the thermal linear expansion coefficients. The $d_0$ parameter for 
$x = 1.5$ case has been chosen to obtain a single representative value of the 
H-bond length. In order to have a non-vanishing current, the electric field 
$K$ must be nonzero, but as long as we stay in the linear response regime, 
which is indeed what takes place in this case, its absolute value has no 
significant effect on the results. The $b$ and $I$ parameters can be derived 
from the geometry of the BnIm, $\nu_R^0$ is calculated from $V_\mathrm{act}$, 
whereas other parameters are treated as free (see \cite{PRE} for details). 
The best fit parameter values for $x = 1.5$ are presented in Table \ref{tab1}.
\begin{table}[thb]
\caption{Values of parameters for the PSSA+BnIm simulations.}
\begin{tabular}{@{}*{3}{l}}
\hline
Parameter  & Symbol  & Value \\
\hline
Frequency of rotation prefactor & $\nu_R^0$        & $1.42\times 10^{12}$ Hz \\
Activation energy for rotations & $V_\mathrm{act}$  & 0.45 eV\\
Rods length                     & $b$              & 3.84 \AA\\
Moment of inertia               & $I$              & 123.6 u \AA$^2$ \\
External electric field         & $K$              & 0.005 $V/$\AA \\
H-bond length                   & $d_0$            & 2.8 \AA \\
Thermal expansion coefficient   & $d_1$            & $10^{-5}$ \AA$/$K \\
$V_a$ barrier height            & $h(T_0)$         & 0.625 eV \\
Distance between minima of $V_a$& $\Delta z(T_0)$  & 0.96 \AA \\
Reference temperature           & $T_0$            & 293.3 K \\
$D$ and $L$ defects energy      & $V_\mathrm{Coul}$ & 0.4 eV \\
Frequency of hopping prefactor  & $\nu_T^0$        & $10^{7}$ Hz \\
Lattice vibration amplitude     & $a_0$            & 0.8 \AA \\
Thermal susceptibility of $a$   & $a_1$            & 0.00093 \AA$/$K \\
\hline
\end{tabular}
\label{tab1}
\end{table}

The variable parameters for the various molar ratios:\\
Fig. \ref{plot2a}:\\
$x=1.0$: $d_0=2.8375$ \AA,\\
$x=0.5$: $d_0=2.83$ \AA.\\
Fig. \ref{plot2b}:\\
$x=1.0$: $d_0=2.8075$ \AA, $V_\mathrm{act}=0.75$ eV, 
$\nu_R^0=1.82~\times~10^{12}$~Hz, \\
$x=0.5$: $d_0=2.815$ \AA, $V_\mathrm{act}=0.675$ eV, 
$\nu_R^0=1.72~\times~10^{12}$~Hz.

\end{document}